\documentstyle[prl,aps,epsfig,prbbib]{revtex}
\begin{document}
\draft
\twocolumn[\hsize\textwidth\columnwidth\hsize\csname  
@twocolumnfalse\endcsname

\title{BCS-BEC crossover at finite temperature for superfluid trapped
Fermi atoms}
\author{A. Perali, P. Pieri, L. Pisani, and G.C. Strinati}
\address{Dipartimento di Fisica, UdR INFM,
Universit\`{a} di Camerino, I-62032 Camerino, Italy}
\maketitle

\begin{abstract}
We consider the BCS-BEC crossover for a system of trapped Fermi atoms at
finite temperature,
both below and above the superfluid critical temperature, by including
fluctuations beyond mean field.
We determine the superfluid critical temperature and the
pair-breaking temperature as functions
of the attractive interaction between Fermi atoms, from the weak- to
the strong-coupling limit
(where bosonic molecules form as bound-fermion pairs).
Density profiles in the trap are also obtained for all temperatures
and couplings.
\end{abstract}
]
\narrowtext

Recent experimental advances with trapped Fermi atoms enable one to reach
considerably lower
temperatures than obtained previously, as well as to vary the effective
attraction between
Fermi atoms via Fano-Feshbach resonances \cite{exp-low-T}.
It then becomes possible to reach conditions where Cooper pairs of Fermi
atoms form in weak coupling
below the superfluid critical temperature $T_{c}$, and composite bosons
form and Bose-Einstein condense
in strong coupling.
It thus appears relevant to formulate a theory of the BCS-BEC
crossover for trapped
Fermi atoms for \emph{all\/} temperatures in the broken-symmetry phase
below $T_{c}$, connecting it
with continuity to the results for the normal phase above $T_{c}$.
This is the main task of the present paper, where a unified theoretical
framework is set up for all
temperatures and couplings.

Limited theoretical results are so far available for the BCS-BEC crossover
in a trap.
A previous study of the density profiles over the whole crossover has
dealt with the
zero-temperature case within a mean-field approach \cite{PPS-RC-03}.
Finite temperatures below $T_{c}$ have been considered  within mean field
for a single coupling value
in the weak-to-intermediate region \cite{Chiofalo}.
Fluctuations over and above mean field have been included over the whole
crossover for temperatures
above $T_{c}$.\cite{Griffin,milstein}

In the present paper, we provide a systematic study of the whole BCS-BEC 
crossover in a trap by including
fluctuations beyond mean field, for all temperatures below and above
$T_{c}$, and up to the pair-breaking
temperature $T^{*}$.
Below the critical temperature, our theory recovers the BCS results in weak
coupling and the
Bogoliubov description for the composite bosons in strong coupling, and
provides an interpolation
scheme in the intermediate (crossover) region where no small parameter
exists for controlling the
approximations of many-body theory.

Study of the BCS-BEC crossover started with the pioneering work by Eagles
for low-carrier doped
superconductors \cite{Eagles}.
A systematic approach to the problem was later given by Leggett
\cite{Leggett}, who showed that a
\emph{smooth\/} crossover from a BCS ground state of overlapping Cooper
pairs to a condensate of
composite bosons occurs as the strength of the fermionic attraction is
increased.
This study was later extended to finite temperatures above $T_{c}$ by
Nozi\`{e}res and Schmitt-Rink
with the use of diagrammatic methods \cite{NSR}.
Extension of this approach to trapped fermions has relied so far mostly on
a local Thomas-Fermi (TF)
approximation \cite{Chiofalo,Griffin,PPS-RC-03}.
This local approximation is also adopted in the present paper.

Our main results for the BCS-BEC crossover in a trap at finite temperature
are the following:

\noindent
(i) We find that the critical temperature $T_{c}$ increases
monotonically from weak to
strong coupling, reaching eventually the value of the Bose-Einstein temperature
for the composite bosons in the trap.
Correspondingly, \emph{no maximum\/} is found in the intermediate-coupling
region for the trapped case.
The presence of this maximum was instead found with the same diagrammatic
theory formulated for the homogeneous
case \cite{NSR}.

\noindent
(ii) We find that in the intermediate-to-strong coupling region the density
profiles show a
characteristic \emph{secondary peak\/} located away from the trap center, 
at temperatures below but close to $T_{c}$.
The occurrence of this peak is due to the combined presence of condensed
and noncondensed composite bosons.
We find that this peak survives up to couplings near the crossover region,
such that the residual
interaction between the composite bosons is strong enough for the peak to
be well pronounced.
In this way, this peak could be experimentally accessible, providing one
with a characteristic feature
of the superfluid state.

\noindent
(iii) We find that the ``pairing fluctuation'' region between $T_{c}$ and 
$T^{*}$, where precursor pairing effects should occur,
is considerably reduced in the trap with respect to the homogeneous case.
Pseudogap phenomena are thus expected to be very much reduced for trapped
Fermi atoms, with respect to
what occurs for high-temperature superconductors \cite{pseudogap-HTCS}.

The system we consider is a gas of Fermi atoms confined in a trap by a
harmonic spherical potential $V(r)$
(where $r$ measures the distance from the trap center).
The Fermi atoms equally populate two spin (hyperfine) states and are
mutually interacting via a point-contact
($s$-wave) attraction.
This attraction is suitably regularized via the scattering length $a_{F}$
of the associated (fermionic) two-body
problem.
The coupling strength is then identified with the dimensionless parameter
$(k_{F}a_{F})^{-1}$, where the
Fermi wave vector $k_{F}$ is related to the Fermi energy $E_{F}=(3N)^{1/3}
\omega$ for noninteracting
fermions in the trap \cite{Pethick-Smith} by $E_{F}=k_{F}^{2}/(2m)$.
Here, $N$ is the total number of Fermi atoms, $\omega$ the trap frequency,
and $m$ the fermion mass (we set
$\hbar = 1$ throughout).
In principle, $(k_{F}a_{F})^{-1} \approx - \, \infty$ corresponds to the
(extreme) weak-coupling and
$(k_{F}a_{F})^{-1} \approx + \, \infty$ to the (extreme) strong-coupling limit.
In practice, the crossover between these limits occurs in the limited interval
$- 1 \lesssim (k_{F}a_{F})^{-1} \lesssim +1$. \cite{footnote_reverse}

The many-body diagrammatic structure for the homogeneous case is
considerably simplified by the use of the
above regularization \cite{Pi-S-98}.
In particular, in the broken-symmetry phase below $T_{c}$ a diagrammatic
theory for the BCS-BEC crossover can be
set up \cite{PPS-long-03} in the spirit of the $t$-matrix 
approximation\cite{footnote-unitary}. This theory includes fluctuation
corrections to the BCS results in weak
coupling and describes the composite bosons in strong coupling by the
Bogoliubov theory\cite{footnoteaB}.
In the present paper, we extend this approach  to the trapped case, by adopting
a \emph{local\/} TF approximation to take into account the trapping potential.
This local approximation is implemented by replacing the chemical potential
$\mu$, whenever it occurs in
the single-particle self-energy and Green's functions, by the local expression
$\mu(r) = \mu - V(r)$.
At the same time, the order parameter $\Delta$ is replaced by a local
function $\Delta(r)$ to be determined
consistently.

Quite generally, in the BCS-BEC crossover approach the chemical potential
is strongly renormalized
when passing from the weak- to the strong-coupling limit.
In our case, the coupled equations for the chemical potential and the local
order parameter $\Delta(r)$ are:
\begin{eqnarray}
\Delta(r) &=&  - \, \frac{4 \pi a_{F}}{m} \, \int \! 
\frac{d{\mathbf k}}{(2\pi)^{3}} \,
\left[ \frac{1}{\beta} \sum_{s} \,
{\mathcal G}_{12}({\mathbf k},\omega_{s};\mu(r),\Delta(r))\right.\nonumber\\
 &-&\left.
\frac{m \, \Delta(r)}{|{\mathbf k}|^{2}} \right]
\label{crossover-local-equation-Delta} \\
N  &=&  8 \pi \int dr \, r^{2} 
\int \frac{d{\mathbf k}}{(2\pi)^{3}} \nonumber \\
&\times & \frac{1}{\beta} \sum_{s}
e^{i\omega_{s}0^+} G_{11}({\mathbf k},\omega_{s}; \mu(r),\Delta(r))
\label{crossover-local-equation-N}
\end{eqnarray}
where ${\mathbf k}$ is a wave vector and $\omega_{s}=(2s+1)\pi/\beta$ ($s$
integer) a fermionic Matsubara frequency
($\beta$ being the inverse temperature).
In the above expressions, 
${\mathcal G}_{12}({\mathbf k},\omega_{s}; \mu(r),\Delta(r))$ 
is obtained from the BCS
anomalous single-particle Green's function 
$\Delta / (E({\mathbf k})^{2} + \omega_{s}^{2})$
(with $E({\mathbf k})=\sqrt{\xi({\mathbf k})^{2}+\Delta^{2}}$ and
$\xi({\mathbf k}) = {\mathbf k}^{2}/(2m) - \mu$),
by replacing $\mu$ with $\mu(r)$ and $\Delta$ with $\Delta(r)$;
the dressed normal single-particle Green's function 
$G_{11}({\bf k},\omega_{s}; \mu(r),\Delta(r))$
is obtained by a similar replacement made on the function 
$G_{11}({\bf k},\omega_{s})$, which contains fluctuation
corrections beyond mean field.

The quantity $G_{11}({\mathbf k},\omega_{s})$ results from the solution of
the $2 \times 2$ Dyson's equation in Nambu notation, with self-energy
\begin{eqnarray}
\Sigma_{11}&(&{\mathbf k},\omega_{s})=- \Sigma_{22}({\mathbf k},-
\omega_{s}) \nonumber\\
=&-& \, \int \! \frac{d {\mathbf k}}{(2\pi)^{3}} \, \frac{1}{\beta} \sum_{\nu}
\, \Gamma_{11}({\mathbf q},\Omega_{\nu}) \, {\mathcal G}_{11}
({\mathbf q}-{\mathbf k},\Omega_{\nu}-\omega_{s})
\label{sigma-11}\\
\Sigma_{12}&(&{\mathbf k},\omega_{s})= \Sigma_{21}({\mathbf k},\omega_{s})
\, = \, - \Delta
\label{sigma-12}
\end{eqnarray}
where $\Omega_{\nu}=2\pi\nu/\beta$ ($\nu$ integer) is a bosonic Matsubara
frequency.
Here, ${\mathcal G}_{11}(k) = - (\xi({\mathbf k}) + i
\omega_{s})/(E({\mathbf k})^{2} + \omega_{s}^{2})$
is the BCS normal single-particle Green's function and
$\Gamma_{11}(q) = \chi_{11}(-q)/[\chi_{11}(q) \chi_{11}(-q) -
\chi_{12}(q)^{2}]$ is the normal pair propagator,
with
\begin{eqnarray}
- \, \chi_{11}(q)&=&  \int \frac{d {\mathbf k}}{(2\pi)^{3}} \, \left[
\frac{1}{\beta} \, \sum_{s} \,
{\mathcal G}_{11}(k+q) \, {\mathcal G}_{11}(-k)\right.\nonumber \\ 
&-&\left. \frac{1}{2 E({\mathbf k})} \right]\label{A-definition} \\
\chi_{12}(q) &=&  \int \! \frac{d {\mathbf k}}{(2\pi)^{3}} \,
\frac{1}{\beta} \, \sum_{s} \,
{\mathcal G}_{12}(k+q) \,{\mathcal G}_{21}(-k)
\label{B-definition}
\end{eqnarray}
and the four-vector notation $k=({\mathbf k},\omega_{s})$ and
$q=({\mathbf q},\Omega_{\nu})$.

Neglecting the diagonal elements (\ref{sigma-11}) of the self-energy
results in the BCS (mean-field)
approximation.
When extrapolated toward strong coupling, this approximation accounts for
the formation of bound-fermion
pairs upon lowering the temperature below $T^{*}$.
Inclusion of the diagonal elements (\ref{sigma-11}) of the self-energy is
required to describe condensation
of these pairs at the lower temperature $T_{c}$.
In strong coupling, the normal
pair propagator $\Gamma_{11}$ (together with
its anomalous counterpart) reduce to the propagators for composite bosons
within the Bogoliubov approximation.
Above $T_{c}$, the diagonal elements (\ref{sigma-11}) correspond to the
$t$-matrix approximation
in the normal phase.

Figure~1 compares the temperature vs coupling phase diagram for the trapped
(t) and homogeneous (h) case,
where $T_{c}$ and $T^{*}$ are identified (and normalized to the 
respective Fermi temperature $T_{F}$ for the two cases).
The temperatures $T_{c}$ and $T^{*}$ are obtained by solving the coupled
equations
(\ref{crossover-local-equation-Delta}) and
(\ref{crossover-local-equation-N}) when $\Delta(r)=0$, with and without
inclusion of the diagonal self-energy (\ref{sigma-11}), in the order.

Note that $T_{c}^{t}$ increases \emph{monotonically\/} from weak to strong
coupling, approaching the value
$T_{BE}=0.94 \omega (N/2)^{1/3}$ of the Bose-Einstein temperature for the
composite bosons in the trap
\cite{Pethick-Smith} (with the same trap frequency for fermions and composite
bosons).
No maximum for $T_{c}$ is thus found in the intermediate-coupling region
for the trapped case, contrary to
the homogeneous case where a maximum occurs at $(k_{F}a_{F})^{-1} \cong 0.35$.
This behavior is consistent with the fact that, for a dilute Bose gas, 
interaction effects lead to a positive (negative) shift of the critical 
temperature in the homogeneous (trapped) case \cite{Baym}. Together with
the vanishing of $T_c$ in weak coupling, this implies that
(at least) one maximum must be present for the homogeneous case, while
the presence of a maximum is not required for the trapped case. It is
encouraging that our approximate theory leads to curves
for the critical temperature in line with these general 
expectations.

Note further that the ``pairing fluctuation'' region of the phase diagram,
delimited in each case by the curves
$T^{*}$ and $T_{c}$, 
\begin{figure}
\centerline{\psfig{figure=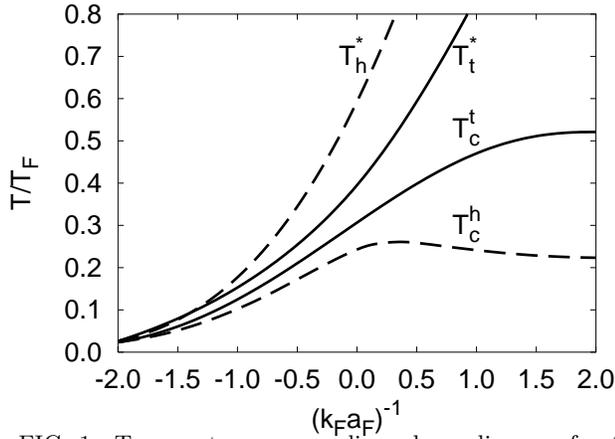,width=6cm,angle=-90}}
\caption{Temperature vs coupling phase diagram for the trapped (full lines)
and homogeneous (dashed lines) case, with the critical 
temperature $T_{c}$ and the
pair-breaking temperature $T^{*}$ shown for the two cases.
Each temperature is normalized to the respective Fermi temperature $T_{F}$.
With this normalization the phase diagram is valid also for anisotropic traps.}
\end{figure}
\noindent
is considerably reduced in the trap with respect to
the homogeneous case.
In the strong-coupling limit, the reduction of the pairing fluctuation region  
stems from the difference in the
density of states $D(\epsilon)$ at energy
$\epsilon$ for noninteracting particles,
when passing from the homogeneous ($D_{h}(\epsilon) \propto
\epsilon^{1/2}$) to the trapped
($D_{t}(\epsilon) \propto \epsilon^{2}$) case.
This difference is, in fact, known to account for the \emph{larger\/} value
of $(T_{BE}/T_{F})_{t}$ with respect
to $(T_{BE}/T_{F})_{h}$ \cite{Pethick-Smith}.
By a similar token, it can be shown that the same difference in the density
of states accounts
for the \emph{smaller\/} value of $(T^{*}/T_{F})_{t}$ with respect to
$(T^{*}/T_{F})_{h}$.

Figure~2 shows the density profiles $n(r)/N$ (such that $4 \pi \int \! dr \,
r^{2} \, n(r)/N = 1$) vs $r/R_{F}$
for three characteristic couplings, from $T=0$ to $T^{*}$ (where
$R_{F}=\sqrt{2E_{F}/(m \omega^{2})}$ is the TF
radius for noninteracting fermions in the trap).
On the weak-coupling side of the crossover ($(k_{F}a_{F})^{-1} =
-1.0$), $n(r)$ depends mildly on
temperature.
In weak coupling, no information can thus be extracted from $n(r)$ about
when the superfluid phase is entered.
For intermediate couplings ($(k_{F}a_{F})^{-1} = 0.0$), $n(r)$ starts to
depend sensibly on
temperature.
Eventually in strong coupling ($(k_{F}a_{F})^{-1} = 1.0$), $n(r)$ shows a
marked temperature dependence.
In this case, the broad density profile at $T^{*}$ corresponds to a system
of noninteracting fermions at the
same temperature.
Upon lowering the temperature, the density profile shrinks considerably and
$n(r=0)$ increases correspondingly,
as expected for a system of weakly-interacting (composite) bosons
\cite{DGPS-99}.

\begin{figure}
\centerline{\psfig{figure=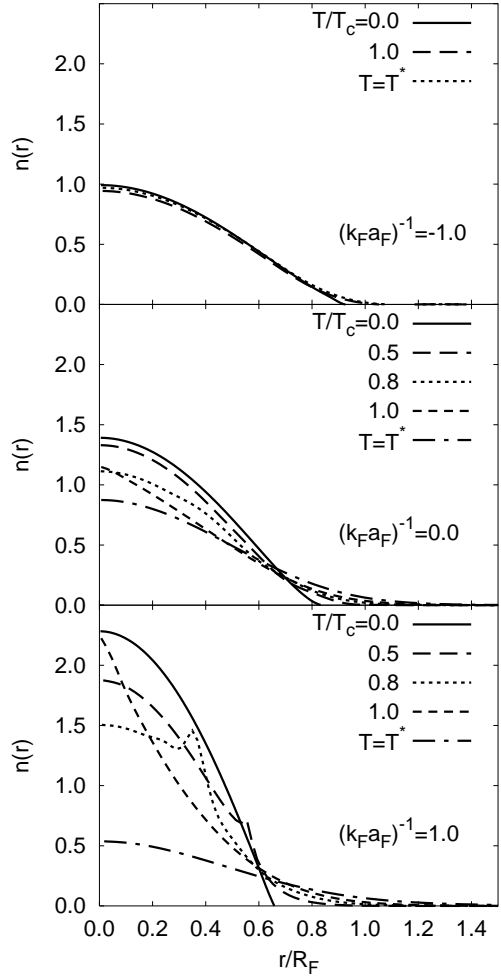,width=12cm,angle=-90}\vspace{1.0cm}}
\caption{Density profile $n(r)/N$ vs $r/R_{F}$ for three coupling
values and several temperatures from $T=0$ to $T^{*}$.}
\end{figure}

Note, in addition, the presence of \emph{a secondary peak\/} away from the
trap center, that shows up at
temperatures below but close to $T_{c}$ (this peak is most evident at about
$0.7 T_{c}$). The presence of this peak stems from the ability of our
theory to recover the Bogoliubov approximation for the composite bosons.
The presence of this peak was, in fact, 
predicted for a weakly-interacting trapped Bose gas \cite{HZG-97}.
Our results show that this peak appears not only in the strong-coupling
limit (which corresponds to weakly-interacting
bosons), but also in the crossover region.
In this region, the residual interaction between the composite bosons is
sufficiently strong for the peak to be
well pronounced over the background, contrary to what occurs for
weakly-interacting bosons.
In this way, the presence of the secondary peak in $n(r)$ below $T_{c}$
could be subject to experimental testing,
providing one with a characteristic signature of the superfluid state.

It is interesting to separate the total density $n(r)$ for a given coupling
into three components, namely,
$n_{F}(r)$ for unbound fermions, $n_{0}(r)$ for condensed pairs,
and $n'(r)$ for noncondensed pairs.
These components are obtained from expressions similar to
(\ref{crossover-local-equation-N}), with $G_{11}$
therein replaced, respectively, by the Green's function ${\mathcal G}_{0}$ for
noninteracting fermions and by the differences 
${\mathcal G}_{11} - {\mathcal G}_{0}$ and ${G}_{11} -
{\mathcal G}_{11}$.
By this procedure we project out from the many-body 
state its fermionic and bosonic character, not only in the extreme weak- and 
strong-coupling regimes where these components have independent 
physical reality, but also in the intermediate-coupling region of interest.

\begin{figure}
\centerline{\psfig{figure=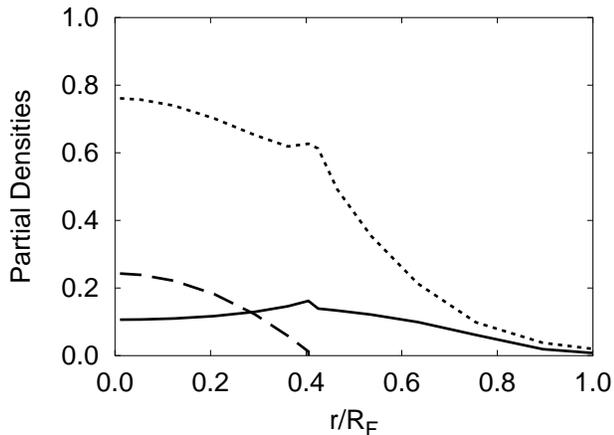,width=6cm,angle=-90}}
\caption{Partial density profiles vs $r/R_{F}$ for $(k_{F}a_{F})^{-1}
= 0.0$ and $T=0.8T_{c}$: $n_{F}(r)/N$ (full line), $n_{0}(r)/N$ 
(dashed line), and $n'(r)/N$ (dotted line).}
\end{figure}

The three components are plotted in Fig.~3 for the coupling value
$(k_{F}a_{F})^{-1} = 0.0$ and the temperature
$0.8T_{c}$.
A layered structure results for these densities, from an inner
core of Bose superfluid to an outer
layer of normal Fermi liquid, in agreement with recent arguments for a
dilute atomic-molecular Fermi cloud
\cite{Bulgac-03}.
Note that, for this intermediate coupling, the component $n'(r)$
obtained by including fluctuations
dominates over the condensate component $n_{0}(r)$ that results 
from mean field.
This implies that, already in the intermediate-coupling region, it is not
possible to rely only on a
mean-field calculation \cite{Chiofalo} to obtain density profiles for
trapped fermions at finite temperature.

In conclusions, with a single theory that includes fluctuations beyond mean
field, both below and above the
critical temperature, we have studied the BCS-BEC crossover for a system of
trapped Fermi atoms at finite
temperature. Novel features, peculiar to the trapped case, have been
contrasted with the results for the homogeneous case.

We are indebted to G. Modugno for discussions.
Financial support from the Italian MIUR under contract COFIN 2001
Prot.2001023848 is gratefully acknowledged.


\begin{thebibliography}{99}
\bibitem{footnote-unitary} The many body 
t-matrix depends on the energy and the 
wave vectors of the incoming particles, as well as on temperature $T$ and 
chemical potential $\mu$. 
In this way, its value does not coincide with the two-body expression 
$4 \pi a/m$ and thus does not diverge when the two-body resonance 
is crossed. 
The t-matrix self-energy is thus "unitarity" limited in the many-body 
sense.
\bibitem{footnoteaB} The Bogoliubov description of the composite bosons is 
here associated with the value $a_B=2 a_F$ for the bosonic scattering length 
$a_B$, which amounts to the Born approximation for the scattering between 
composite bosons\cite{Pi-S-98}. A complete treatment of the composite boson
scattering problem leads to a reduction of $a_B$ by about a factor of 
3\cite{Pi-S-98,petrov}. Owing to the very slow dependence of the density
profile on $a_B$ in strong-coupling, we expect this
difference to make no significant change in this quantity.
\end{thebibliography}
\end{document}